\documentclass[10pt, a4paper]{article}       

\usepackage{comment}
\usepackage{microtype}

\usepackage{graphicx}

\usepackage{mathptmx}      
\usepackage[T1]{fontenc}
\usepackage{graphics,float,amsmath,amstext,lscape,amssymb}
\usepackage{color}
\usepackage{multirow}
\usepackage{adjustbox}

\usepackage{tikz}
\usepackage{hyperref}

\usepackage[affil-it]{authblk}



\providecommand{\keywords}[1]{\\ \\ \textbf{\textit{Keywords: }} #1}
\providecommand{\subclass}[1]{\\ \\ \textbf{\textit{Subject Classification: }} #1}

\begin{document}
	
	\title{Latent likelihood ratio tests for assessing spatial kernels in epidemic models
	}
	
	
	\author[1]{David Thong}
	\author[1]{George Streftaris}
	\author[1]{Gavin J. Gibson}
	\affil[1]{Maxwell Institute for Mathematical Sciences, Heriot-Watt University, Riccarton, Edinburgh, EH14 4AS}

\maketitle

\begin{abstract}
One of the most important issues in the critical assessment of spatio-temporal 
stochastic models for epidemics is the selection of the transmission kernel used to represent the 
relationship between infectious challenge and spatial separation of infected and susceptible hosts. 
As the design of control strategies is often based on an assessment of the distance over which transmission 
can realistically occur and estimation of this distance is very sensitive to the choice of kernel function, 
it is important that models used to inform control strategies can be scrutinised in the light of observation 
in order to elicit possible evidence against the selected kernel function.  While a range of approaches to model criticism 
are in existence, the field remains one in which the need for further research is recognised.  In this paper, 
building on earlier contributions by the authors, we introduce a new approach to assessing the validity of 
spatial kernels - the latent likelihood ratio tests - and compare its capacity to detect model misspecification 
with that of tests based on the use of infection-link residuals.  We demonstrate 
that the new approach, which combines Bayesian and frequentist ideas by treating the statistical decision maker 
as a complex entity, can be used to formulate tests with greater power than infection-link residuals to 
detect kernel misspecification particularly when the degree of misspecification is modest. This new approach avoids the use of a fully Bayesian approach which may introduce undesirable complications related to computational complexity and prior sensitivity. 
\subclass{62F15,   92-08,   92D30,   62M30}
\end{abstract}

\section{Introduction \label{intro}}
Selection of spatial kernel functions in spatio-temporal epidemic models is a question of paramount practical importance.  
It is recognised \cite{shaw93,gibson96} that predictions regarding the speed of epidemic spread or propensity for transmission 
over long distances are very sensitive to the choice of spatial kernel function.
The control of epidemics such as foot and mouth disease (FMD) in the UK
\cite{Keeling2001,batesfmdarticle,chis2009epidemiological,Ferguson2001,Ferguson2001a,Jewell2009,Morris2001,bbcfmdtimeline,ChisSter2007,streft04,Tildesley2008}
or citrus canker in the USA \cite{neri2014bayesian,Gottwald2002a,Gottwald2002} 
has proved controversial on account of the removal of healthy hosts as part of the strategy.  
Such strategies have been informed by mathematical models in which the choice of spatial kernel 
has been a factor in determining a `culling radius' (for example \cite{Keeling2001,Ferguson2001}).  Methods for model criticism and comparison are therefore much-needed to 
ensure that, as far as possible, such decisions can be supported and defended in the light of available evidence.  

Although several approaches to model criticism for epidemic models exist, in the epidemic
context many of these suffer from certain difficulties which motivate the development of further approaches.    
In \cite{Gibson17} the approaches commonly used are reviewed.  These range from Bayes factors 
and Bayesian model selection, posterior predictive p-values, latent classical tests and the use of 
the DIC including missing data variants. One recommendation from \cite{Gibson17} is that 
it is prudent to follow the advice of Box  \cite{Box} that one should test selectively for those forms 
of misspecification which are most strongly suspected and to design specific tests for this purpose. 
This is the approach that is taken throughout this paper where we will formulate latent likelihood 
ratio-tests \cite{streft04,streft12} for kernel misspecification and compare their sensitivity with that of the infection-link 
residuals test introduced in \cite{Lau14}.  Both of these methods are examples of latent classical testing, 
an approach which fuses Bayesian and classical thinking by having a Bayesian observer impute the result of a 
classical goodness-of-fit test applied to a latent process, where the process and the test can be specified flexibly 
to maximise the chance of detecting the suspected misspecification, should it be present. The approach differs from a purely
Bayesian one, in which modes of misspecification are accommodated through the process of Bayesian model expansion.
One reason for not adopting this latter approach is that inference for relatively simple epidemic models using partial observation
is already a complex process. We therefore seek model comparison methods that can be utilised without increasing the dimension
of the models to which Bayesian methods are applied. Accordingly, the methods we present can be integrated into analyses without 
increasing the complexity of the fundamental Bayesian computations.

We will consider stochastic models for an infectious disease spreading through a closed population of spatially-distributed
hosts exemplified by the spatio-temporal Susceptible-Exposed-Infectious-Removed (SEIR) model.
It will be assumed that the locations of hosts are known and fixed. Under this model, the host population at 
time $t$ is partitioned into subsets $S(t)$, $E(t)$, $I(t)$ and $R(t)$.  Hosts in $S(.)$ are susceptible to
infection, hosts in $E(.)$ have been infected but are not yet able to transmit, hosts in $I(.)$ can pass on infection,
while hosts in $R(t)$ have been removed (e.g. by death, hospitalisation, or the acquisition of immunity) and play no further part in 
the epidemic.  A susceptible individual at coordinates $\bf x$ at time $t$ becomes exposed at a rate 
\begin{equation}
\label{rate}
R(t) = \alpha + \beta \sum_{{\bf y} \in I(t)} K(\kappa, {\bf x}, {\bf y})
\end{equation}
where $I(t)$ comprises sites infectious at time $t$, $\alpha$ and
$\beta$ are {\em primary} and {\em secondary} infection rates, and $\kappa$ parametrises 
the {\em spatial kernel function} $K()$. For convenience, we identify hosts with their location. The choice of $K$ greatly influences the design
of control strategies, for example based on ring-culling. A longer-tailed kernel may suggest the use of a larger culling radius and vice versa. Sojourn times in the $E$ and $I$ class are modelled using 
appropriate distributions such as Gamma or Weibull distributions. We will denote by $\theta$ the vector of
model parameters formed from $\alpha$, $\beta$, $\kappa$ supplemented by parameters 
specifying the distributions of sojourn times in $E$ and $I$.  This flexible framework can accommodate
complexity arising e.g. from host heterogeneity as appropriate \cite{parry14,jewell}. 

When data $y$ contain partial information (e.g. removals or `snapshots' of $I(t)$ using imperfect
diagnostic tests) data-augmented Bayesian analysis is now a standard tool for investigating $\pi(\theta|y)$
via $\pi(\theta, z |y)$, where $z$ incorporates unobserved transitions and, possibly,
graphs of infectious contacts. Computations are often effected using reversible-jump MCMC 
or particle filtering \cite{king08}. In this paper, we will assume that observations include times and locations of all transitions from E to I
and from I to R, so that the subsets $I(t)$ and $R(t)$ are observed but individuals in $S(t)$ cannot be distinguished 
from those in $E(t)$. We therefore specify $z$ to incorporate the times and location of the unobserved transitions from S to E (termed {\em exposure}
events) and use MCMC to sample from $\pi(\theta, z |y)$.  As the number of exposure events is not uniquely determined by the data, the state-space for 
$(\theta, z)$ comprises components of varying dimension requiring the use of reversible-jump methods. It is straightforward to apply the methods used on this class of models, to snapshot data.

The rest of the paper is organised as follows.  In Section 2, we 
discuss the general features of the latent classical testing framework 
before describing how functional-model representations of epidemic models 
have been used in the specification of {\em infection-link residuals} \cite{Lau14}.  In Section 3,
we explicitly formulate new latent classical tests for detecting kernel misspecification using likelihood ratios, where the ratio is
based on a complete or partial parameter likelihood.  In Section 4, we apply the tests to simulated data comparing
the ability or `power' of the likelihood-based and infection-link residual tests to detect kernel misspecification 
in several scenarios.  Conclusions are summarised in Section 5.

\section{Latent classical testing and residual construction}

Throughout we consider the situation where a Bayesian observer $B$ observes the outcome $y$ of an 
experiment for which they have proposed a statistical model $\pi_0(y | \theta)$ where beliefs regarding 
the parameter vector $\theta$ are represented by the prior distribution $\pi_0(\theta)$. 
We suppose that the likelihood $\pi_0(y | \theta)$ may not necessarily be tractable - a situation 
which typically applies in the case of a partially observed epidemic. Now let $r$ be some process 
varying jointly with $y$ and suppose that we have a model $\pi(y, r | \theta)$ for which the marginal 
model $\pi(y|\theta)$ coincides with $\pi_0(y | \theta)$.  Suppose that the model $\pi(r | \theta)$ 
is tractable. Then in the latent classical framework, the model $\pi_0(y | \theta)$ is 
assessed by having $B$ impute the result of a classical test of the model $\pi(r | \theta)$ which 
is carried out by a classical observer $C$ of $r$.  The evidence found by $C$ against $\pi(r |\theta)$, 
for example as summarised by a P--value $p(r; \theta)$, can be considered as evidence against the joint model $\pi(y, r| \theta)$.

In \cite{Gibson17} it is discussed how the roles of $B$ and $C$ above are analogous to those of 
the Freudian ego and superego, with model formulation and fitting being done by the former in the 
Bayesian framework and model criticism by the latter in the classical framework. The conclusions of 
the analysis carried out in this `dual-observer' framework are necessarily presented via $B's$ posterior distribution of $C's$ P--value, $\pi(p(r, \theta | y))$ from 
which $B$ can extract natural measures of lack of fit such as $\Pr(p(r, \theta) < \alpha | y)$ for some
suitably small $\alpha$. Note that this approach can be viewed as an extension of the framework
of posterior predictive checking.  The main differences lie in the use of latent processes 
to specify the p-value and in the consideration of the entire distribution of p-values as opposed 
to its mean as captured by a posterior predictive p-value \cite{meng}.
As noted by Meng, `... every problem is a missing data problem...' and the approach we take exploits this.  
Suppose that $\pi_j(y, r_j |\theta), j = 1, ..., k$ represent 
models for the joint distribution of $(y, r_j)$ all of which specify the 
same {\em marginal} model $\pi_0(y|\theta)$ and share 
a common parameter prior distribution $\pi(\theta)$.  Then observation of $y$ alone
carries no information on the relative validity of these models. 
That is, $y$ carries exactly the same 
evidence {\em against} every model with marginal $\pi_0(y | \theta)$. 
Therefore, the latent process $r$ can be designed to yield
a test tailored to detecting the suspected form of misspecification.
This facility is exemplified by the construction of infection-link residuals \cite{Lau14}.

\subsection{Infection-link residuals}

The starting point is to construct a functional-model representation of the epidemic process.  In this 
formalism the observations $y$ are represented as a deterministic function 
$x = h(r, \theta)$ of $\theta$ and some unobserved process $r$ with fixed distribution independent of
$\theta$, where $x=(y,z)$. This means that $r$ can be treated as a residual process and tests for compliance with the 
specified distribution can be applied to the imputed realisations of $r$. Such an approach fits well for 
epidemic models where sampling from $\pi(\theta, r | y)$ is often possible using 
Markov chain Monte Carlo methods.  

\begin{figure}
\begin{centering}
\begin{tikzpicture}[domain=0:4]
\draw[->] (-0.2,0) -- (4.2,0) node[right] {$d$};
\draw[->] (0,-1.2) -- (0,1.2) node[above] {$f(d) $};
\draw[color=blue] plot (\x,{exp(-\x)}) node[below] {$f(d) = K_{fitted}(\kappa,d)$};
\draw[color=red] plot (\x,{1/(1+\x)}) node[above] {$f(d) = K_{actual}(\kappa,d)$};
\end{tikzpicture}

\end{centering}
\caption{Diagram of motivation for the infection link residual (ILR) $\tilde{r}_{2k}$\label{fig:ILR-motivation}}
\end{figure}
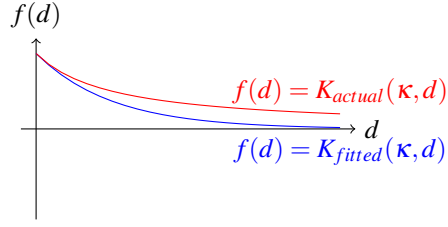

In \cite{Lau14} a functional-model for a spatio-temporal 
SEIR model is presented where the process $r$ is composed of 
four independent i.i.d. U(0, 1) sequences, $r_1, r_2, r_3, r_4$. Consider the mapping $x = h_\theta(r_1, r_2, r_3, r_4)$, where $x$ 
records the time and nature of every event occurring during the epidemic.  Details can be found in
\cite{Lau14}.
The time of each subsequent infection
event is determined from the process $r_1 = \{ r_{1j}, j \geq 1\}$ while processes $r_3$ and $r_4$ specify the quantiles 
of the sojourn periods in the E and I class 
respectively for each infection.  The infection-link residual sequence (to which tests are applied)
$r_2 = \{r_{2j}, j \geq 1 \}$ determine the particular I-S pair responsible for each
infection event. Given the time of the $j^{th}$ infection, $t_j$, 
we identify the set of I-S links
$$
S = \{ K( {\bf x}, {\bf y}, \kappa)  | {\bf x} \in S(t), {\bf y} \in I(t) \}
$$
and order these according to ascending order of magnitude. 
The particular link causing the $j^{th}$ infection 
is selected by considering the cumulative sum of the ordered links and identifying the
first link where this cumulative sum exceeds the value $r_{2j}W$ where $W$ denotes the sum of the weights in $S$. It is straightforward to explore the joint posterior $\pi(\theta, r_1, r_2, r_3, r_4 | y)$. If the kernel function $K$ has been
misspecified (for example by underestimating the propensity for long-range transmission by assuming an exponentially bounded form
when a power-law relation is more appropriate, see Fig. \ref{fig:ILR-motivation}), then when the process $r_2$ is imputed, some systematic deviation from a $U(0, 1)$
should be anticipated. In \cite{Lau14} p-values were imputed from an Anderson-Darling 
test \cite{Anderson1954} applied to $r_2$ and it was demonstrated that
the approach can detect kernel misspecification in simulated data sets.  In this paper 
we investigate whether it is possible to improve on the sensitivity of the ILR tests using likelihood-based methods.

\section{Latent likelihood ratio tests for model comparison \label{sec:llrttestsformodcompar}}

This general approach has been followed previously in \cite{streft12} where 
results of an ANOVA test applied to viraemic measurements taken on a host population, partitioned by depth in
an unobserved infection graph, were imputed. While the ILR test 
is targeted at generic forms of model inadequacy (misspecification of the tail properties
of a spatial kernel), latent likelihood ratio tests demand that a more specific alternative
model is identified.

To test the validity of model $M_0$ with likelihood $\pi_0(y|\theta)$ against a simple alternative model $M_1$ (i.e. with no free parameters) $\pi_1(y)$,  it is natural
for $B$ to impute $C's$ conclusion from a classical test based on a likelihood ratio 
$T(y, \theta) = \frac{\pi_0(y | \theta)}{\pi_1(y)}$.  For epidemic models and data,
$\pi(y |\theta)$ and $\pi_1(y)$ would typically be intractable. Nevertheless, $B$ can impute $(\theta, x)$, 
where $x$ represents an appropriate latent process, and the conclusion of $C's$ test based on
a likelihood ratio $T(x, \theta) = \frac{\pi_0(x | \theta)}{\pi_1(x)}$, 
so long as $\pi_0(x | \theta)$ and $\pi_1(x)$ are tractable. 
It is straightforward to extend the idea to a generalised likelihood ratio test (GLRT) 
when the alternative model is composite by replacing 
$\pi_1(x)$ with $\pi_1(x|\hat \theta_1)$ where $\hat \theta_1$
is the maximum likelihood estimate (MLE) of the parameter 
$\theta_1$ in the alternative model.

\vspace{5mm}
Suppose that, given partial information $y$, we use data-augmented MCMC to explore $\pi_0(\theta, x |y)$ where
$x$ comprises the times and nature of all transitions of all transitions within an observation window
$(0, T_{max})$.  The latent likelihood ratio test may be implemented as an
addendum to this analysis as follows. 

\begin{enumerate}
\item
$B$ draws samples $(\theta, x)$ from $\pi_0(\theta, x | y)$.
\item \label{itm:pevalcalcstep}
For each sample $(\theta^{(i)}, x^{(i)})$:
\begin{itemize}
\item
 $B$ then calculates 
the MLE, $\hat \theta_1^{(i)}$, of the parameter $\theta_1$, 
under the alternative model.
\item
$B$ computes the ratio $T(x^{(i)}, \theta^{(i)}) = \frac{\pi_0(x^{(i)}|\theta^{(i)})}{\pi_1(x^{(i)} | \hat \theta_1^{(i)})}$ and the associated
p-value
$$
p(\theta^{(i)}, x^{(i)}) = \Pr(T(x, \theta^{(i)}) < T(x^{(i)}, \theta^{(i)})  | \theta^{(i)},x^{(i)}),
$$
where $x$ is drawn randomly from $\pi_0(x | \theta^{(i)})$.
\end{itemize}
\end{enumerate}
By repeating these steps within a standard MCMC analysis, 
a sample from $\pi(p(\theta, x) | y)$ can be obtained.  

\vspace{5mm}
Note that we do not assume nesting 
of models that might allow asymptotic results on sampling distributions
of likelihood ratios to be applied. For each sampled pair $(\theta^{(i)}, x^{(i)})$ we may estimate the p-value by
simulation. 
The simplest approach is to estimate the posterior expectation of the p-values as follows:
\begin{itemize}
\item
Compute the ratio  $T(x^{(i)}, \theta^{(i)}) = \frac{\pi_0(x^{(i)}|\theta^{(i)})}{\pi_1(x^{(i)} | \hat \theta_1^{(i)})}$.  Simulate a random draw, $x'$ from
$\pi_0(x|\theta^{(i)})$, obtain the MLE, $\hat \theta'_1$, by maximising $\pi_1(x'|\theta_1)$, and compute 
$T' = \frac{\pi_0(x'|\theta)}{\pi_1(x' |\hat \theta'_1)}$.  
\end{itemize}
An estimate of the posterior mean of $\pi(p(\theta, x) | y)$,
is obtained from the frequency with which $T' < T(x^{(i)}, \theta^{(i)})$. 
This quantity provides some information on the strength 
of the evidence against the modelling assumptions. 

\subsection{Imputation and reinforcement}

The above test may appear to have the potential to be targeted at specific forms of misspecification
but some caveats should be noted. When $B$ imputes the latent process $x$
in order to specify a tractable classical test, they appeal to the
modelling assumptions underlying $M_0$. It follows
that imputation will {\em reinforce} these assumptions to an extent dependent on the
amount of imputed information.  For example, 
if the imputed $x$ included not only unobserved quantities $x_1$ 
from the present experiment
but also a further $m-1$ independent replicates $x_2, ..., x_m$ then, for large
$m$, $\pi(p(\theta, x)| y) \approx {\rm U}(0, 1)$
for a large range of tests, since the test result would be increasingly dominated by the imputed
replicates.

\vspace{5mm}
To understand the impact of imputation more formally, 
we consider the simple situation where $B's$ prior distribution $\pi_0(\theta)$ places 
all belief on a single value $\theta_0$, giving a density $\pi_0(x)$ 
for the latent process $x$.  We assume that the alternative model, $M_1$, 
is simple (i.e. has no free parameters) with sampling distribution $\pi_1(x)$.  
Suppose now that $B$ observes $y = f(x)$, so that $x$ is an augmented version of $y$ and
imputes $x$ via $\pi_0( x | y)$.  They then
impute the p-value, $p_x$ computed by $C$ from an LRT applied to $x$.  Suppose that 
$B$ summarises their posterior belief regarding $C's$ evidence against $\pi_0$ by the quantity
\begin{displaymath}
\gamma_{x,\alpha}(y) = \pi_0(p_x < \alpha | y)
\end{displaymath}
for some suitably small $\alpha$.  A natural analogue of {\em power} for $B$ would be
the expectation of $\gamma_{x,\alpha}(y)$ under the alternative hypothesis, that is
\begin{displaymath}
\beta_x = {\rm E}
\left[ \gamma_{x,\alpha}(y) | {\rm M}_1 \right]
.
\end{displaymath}
Note that when $x \equiv y$ the quantity $\gamma_{y,\alpha}(y)$ is an indicator function 
and $\beta_y$ is the power of the uniformly most powerful test obtained using the Neyman-Pearson Lemma.  
Then we have following result.

\vspace{5mm}
\noindent
{\bf Proposition 3.1.}   {\em For $x$, $y$, ${\rm M}_0$, ${\rm M}_1$ as described above, $\beta_x \leq \beta_y$. }

\vspace{5mm}
\noindent
{\em Proof. }  The most powerful classical 
test of level $\alpha$ of $M_0$ v $M_1$ that can be applied to the imputed $x$ is based on the
ratio $\frac{\pi'_0(x)}{\pi'_1(x)}$ where $\pi'_0$ and $\pi'_1$ represent 
the sampling densities of the {\em imputed} $x$ respectively under $M_0$ and $M_1$.  
Now $\pi'_0(x) = \pi_0(y)\pi_0(x|y) = \pi_0(x)$ 
while $ \pi_1(x) =  \pi_1(y)\pi_0(x|y)$, so that
\begin{displaymath}
\frac{\pi'_0(x)}{\pi'_1(x)} = \frac{\pi_0(y)}{\pi_1(y)}.
\end{displaymath}
Therefore, a LRT applied directly to $y$ is equivalent to a LRT 
applied to $x$ when $x \sim \pi'_0(x)$ and $x \sim \pi'_1(x)$ 
are used as the sampling densities of $x$ under the competing hypotheses. 
We denote by $p_y$ the resulting p-value.

\vspace{5mm}
Now, for the latent likelihood ratio test, $B$ imputes the result of $C's$ 
likelihood ratio test applied to the 
imputed $x$, where $C's$ test is based on the test statistic
\begin{equation}
\frac{\pi_0(x)}{\pi_1(x)} = \frac{\pi'_0(x)}{\pi_1(x)}
\end{equation}
with the associated p-value, $p_x$.  
By the Neyman-Pearson Lemma, the power of this test cannot exceed that of the optimal test.  
We therefore have that for any given value $\alpha$,
\begin{displaymath}
\Pr(p_x < \alpha | {\rm M}_1) \leq \Pr(p_y < \alpha | {\rm M}_1) = \beta_y
\end{displaymath}
where $x \sim \pi'_1(x)$ and $y \sim \pi_1(y)$ on the left and right-hand sides respectively.
Note that
\begin{displaymath}
Pr(p_x < \alpha | {\rm M}_1) = \int \pi_0(p_x < \alpha | y) \pi_1(y)dy = \beta_x.
\end{displaymath}
This completes the proof.

\vspace{5mm}
Now suppose more generally that $M_0$ uses an arbitrary prior $\pi_0(\theta)$, 
while $M_1$ remains simple, and define  $\beta_x(\theta)$ and $\beta_y(\theta)$
in the obvious way. Then under the prior distribution, $\beta_y(\theta)$ is absolutely dominant
over $\beta_x(\theta)$ so that $B$ views with certainty the LRT applied directly to $y$ as giving the more
powerful test of $M_0$ against $M_1$.

\vspace{5mm}
In the above proof the inequality  $\beta_x \leq \beta_y$ arises from the 
disparity between $\pi'_1(x)$ and $\pi_1(x)$. 
We show that this disparity, as characterised using Kullback-Leibler (KL) divergence, increases as the amount of imputation grows.  
Suppose that $y = f(x)$ and $x = g(z)$, so that $z$
represents the outcome of an experiment that is even more informative than $x$. 
Consider again the case of simple hypotheses for $M_0$ and $M_1$,
with $\pi_0$ and $\pi_1$ denoting the respective sampling densities of quantities.

\vspace{5mm}
If the test is based on the imputed $z$ then the optimal test statistic uses
the distribution of this imputed $z$ and is therefore the ratio $\frac{\pi_0(z)}{\pi_1(y)\pi_0(z|y)}$.  
We use $\pi_0^i$ and $\pi_1^i$ 
to denote the sampling densities
of imputed quantities under the respective hypotheses. 
Note that $\pi_0^i = \pi_0$. We now consider the Kullback-Leibler divergence 
between $\pi_1^i(z)$ and $\pi_1(z)$.  This can be calculated as
\begin{eqnarray*}
KL(\pi_1^i, \pi_1) &= &\int \pi_1^i(z) \log (\frac{\pi_1^i(z)}{\pi_1(z)}) dz\\
&= &\int \pi_1^i(z) \log (\frac{\pi_1(y)\pi_0(x|y)\pi_0(z|x)}{\pi_1(y)\pi_1(x|y)\pi_1(z|x)}) dz\\
&= &\int \pi_1^i(z) (\log (\frac{\pi_0(x|y)}{\pi_1(x|y)}) + \log(\frac{\pi_0(z|x)}{\pi_1(z|x)}))dz\\
&= &\int \pi_1^i(z) \log (\frac{\pi_0(x|y)}{\pi_1(x|y)})dz + \int \pi_1^i(z) \log (\frac{\pi_0(z|x)}{\pi_1(z|x)})dz\\
&= &\int \pi_1(y)\pi_0(x|y)\pi_0(z|x)\log (\frac{\pi_1(y)\pi_0(x|y)\pi_0(z|x)}{\pi_1(y)\pi_1(x|y)\pi_0(z|x)})dz\\
&&+ \int \pi_1(y)\pi_0(x|y)\pi_0(z|x) \log (\frac{\pi_1(y)\pi_0(x|y)\pi_0(z|x)}{\pi_1(y)\pi_0(x|y)\pi_1(z|x)})dz\\
\end{eqnarray*}

\vspace{5mm}
\sloppy The first integral above is the KL divergence between the density
$\pi_1^i(z) = \pi_1(y)\pi_0(x|y)\pi_0(z|x)$ and the density $\pi_1(y)\pi_1(x|y)\pi_0(z|x)$. 
Suppose that the latter is used on the denominator in a ratio test statistic applied to the imputed $z$.  
Then this ratio
is clearly $\frac{\pi_0(x)}{\pi_1(x)}$ where $x$ is the imputed value and the power of the test corresponds 
to that of a latent likelihood ratio test applied to $x$.  The second integral above is 
itself a KL divergence greater than zero.  It follows that 
\begin{displaymath}
KL(\pi_1^i(z), \pi_1(z)) > KL(\pi_1^i(z), \pi_1(y)\pi_1(x|y)\pi_0(z|x)).
\end{displaymath}

\fussy In the light of this increasing divergence, we may suspect that the power of a LRT that uses $\pi_1(z)$ on the denominator
may be less than that of a test using $\pi_1(y)\pi_1(x|y)\pi_0(z|x)$ or, equivalently, a latent likelihood ratio test applied
directly to the imputed $x$. When seeking a suitable latent process $x$, 
it may be prudent to minimise the extent of imputation and, consequently, the 
degree of reinforcement of the model under test.
That is, if $y$ is specified by $x$ which, in turn, is specified by $z$, 
then, assuming the likelihoods $\pi_0(x| \theta)$ and 
$\pi_1(x | \theta)$ are tractable, $x$ should be preferred to $z$ as the choice for the 
latent process.

\section{Latent likelihood tests for kernel assessment}

We now return to the situation where we wish to assess the validity of the choice of transmission kernel for a 
spatio-temporal SEIR model for an emerging epidemic based on partial data $y$. 
We construct latent likelihood ratio tests and compare their ability 
with that of the ILR test.

We make the following assumptions. Bayesian observer $B$ proposes an SEIR model for an emerging epidemic of the 
form described in Section \ref{intro}. The model, $M_0$, incorporates a transmission kernel $K_0(d, \kappa_0)$ 
and a prior $\pi_0(\theta)$ is assigned to the parameter vector $\theta_0 = (\alpha, \beta, \kappa_0, \theta_E, \theta_I)$. 
Observer $C$ criticises this model, suspecting an alternative transmission kernel $K_1(d, \kappa_1)$ 
may be more appropriate. All other aspects of the alternative model $M_1$ coincide with $M_0$. 
We denote the parameter in $M_1$ by $\theta_1 = (\alpha, \beta, \kappa_1, \theta_E, \theta_I)$. 
Note that since $M_1$ will be treated by Observer $C$ in the framework 
using frequentist methods, then no prior for $\theta_1$ need be specified.

We consider two forms of latent likelihood test, based on full and partial likelihood 
respectively, which differ in terms of the amount of information imputed for the test.

\subsection{Full-trajectory LLRT}

This analysis is achieved through $B$ investigating $\pi_0(\theta_0, x | y)$, where $x$ is the complete trajectory 
of the epidemic (the waiting times and locations of the exposure, infection and removal events not considering the infection tree). The MCMC algorithm used to do this is standard (for example, \cite{gibsonrenshaw1998,o1999bayesian,Streftaris2004,streft04,Forrester2006,gibson2006bayesian,chis2009epidemiological,Starr2009,neri2014bayesian}) and is summarised in Electronic Appendix 1.
For each sample $(\theta_0, x)$, the MLE $\hat \theta_1$ is computed using the optimisation routine 
described in Electronic Appendix 2, and the algorithm is implemented as in Section \ref{sec:llrttestsformodcompar}. The test statistic used is the full likelihood ratio, as detailed in Step \ref{itm:pevalcalcstep} in Section \ref{sec:llrttestsformodcompar}.

\subsection{Partial LLRT}

In this setting, Observer $B$ investigates $\pi_0(\theta_0, x | y)$ but Observer $C$ is
provided only with $\theta_0$ and $z$, where $z$ incorporates for each exposure event, $j$:
\begin{itemize}
\item
the sets of locations of susceptible and infectious individuals, $S(t_j-)$, $I(t_j-)$ immediately prior to the time of the event, $t_j$;
\item
the location of the exposed individual, ${\bf x}_j \in S(t_j-)$.
\end{itemize}
The times or even the order of the exposure events are not 
included in $z$ though some restrictions on the latter will follow from $z$. 
Let $G_0(\theta_0, z)$ be defined by
\begin{displaymath}
G_0(\theta_0, z) = \prod_j \frac{\alpha + \beta\sum_{{\bf} y \in I(t_j-)} K_0(|{\bf y} - {\bf x}_j|, \kappa_0)}
{|S(t_j-)|\alpha + \beta\sum_{{\bf y} \in I(t_j-), {\bf x} \in S(t_j-)} K_0(|{\bf y} - {\bf x}|, \kappa_0)}
\end{displaymath}
where $|S(t_j-)|$ denotes the cardinality of $S(t_j-)$.
An analogous partial likelihood for $M_1$ with kernel function $K_1$ and parameter $\theta_1$ is given by
\begin{displaymath}
G_1(\theta_1, z) = \prod_j \frac{\alpha + \beta\sum_{{\bf} y \in I(t_j-)} K_1(|{\bf y} - {\bf x}_j|, \kappa_1)}
{|S(t_j-)|\alpha + \beta\sum_{{\bf y} \in I(t_j-), {\bf x} \in S(t_j-)} K_1(|{\bf y} - {\bf x}|, \kappa_1)}
\end{displaymath}
Then, if $\hat \theta_1$ maximises $G_1(\theta_1, z)$ we can define a partial likelihood ratio statistic
$$
T_{partial}(\theta_0, z) = \frac{G_0(\theta_0, z)}{G_1(\hat \theta_1, z)}.
$$
This statistic is used in place of the full likelihood ratio in Step \ref{itm:pevalcalcstep} in Section \ref{sec:llrttestsformodcompar}.

We may motivate the partial LLRT from the perspective of reinforcement.  
The partial LLRT requires that only $\theta_0$ and $z$ are imputed by $B$ for its calculation. 
Thus, the impact of reinforcement of $M_0$ may be lessened.  Moreover, if detection of a possibly 
misspecified kernel is the goal, then $T_{partial}(\theta_0, z)$ is a statistic which `focuses' 
on this aspect of the model. It is therefore possible that the partial LLRT,
at least in some circumstances, may be more effective 
in eliciting evidence of a misspecified kernel than the full-likelihood LLRT.
Moreover, the partial LLRT is a natural comparator for the ILR test used in \cite{Lau14}, as both tests utilise the same information. For both of these tests, 
$(\theta_0, z)$ is necessary and sufficient for computation of the test result.  

In the next section we 
consider the ability of the ILR, and the two LLRTs to detect misspecification 
of the transmission kernel in a spatio-temporal epidemic model in a simulation study.

\section{Simulation study \label{sec:sim-study}}

In keeping with the assumptions of \cite{Lau14},
we assume that the observations $y$ record the transitions from E to I and from I to R, but that exposure 
events are not recorded. Epidemics are simulated in an initially totally susceptible population uniformly 
distributed over a square region of size $2000\times 2000$ units.  Both primary and secondary infection are present 
and an exponentially decaying spatial kernel function of the form
$$
K(\mathbf{x}, \mathbf{y}, \kappa) = \exp(-\theta|\mathbf{x}- \mathbf{y}|)
$$
is assumed, where $\mathbf{x}$ and $\mathbf{y}$ denote the positions of two hosts.
We assume that the sojourn times in the E and I classes follow Gamma distributions with means and variances 
$\mu_E, \mu_I$ and $\sigma^2_E, \sigma^2_I$ respectively. Table \ref{tab:simdataparamentersfordatasets} lists the parameter values used to simulate 
the data. These parameters are based on those used in the simulation study of the ILR test in \cite{Lau14}, to allow comparison with the simulation study therein. Starting from an entirely susceptible population, the epidemic is simulated until complete infection of the population. Four different parameter sets are used - a baseline scenario, and the same parameter set with $\alpha$, $\beta$ and $\kappa$ respectively increased to twice the baseline value.  The baseline 
set of parameter values, and the modified values, are given in Table \ref{tab:simdataparamentersfordatasets}. 

\begin{table}
	
\begin{center}
\begin{tabular}{|c|c|c|c|c|}
\hline 
\multirow{2}{*}{Parameter} & \multicolumn{4}{c|}{Data-set}\tabularnewline
\cline{2-5} 
 & Original & $\alpha\times2$ & $\beta\times2$ & $\kappa\times2$\tabularnewline
\hline 
$\alpha$ & $0.001$ & $\mathbf{0.002}$ & $0.001$ & $0.001$\tabularnewline
\hline 
$\beta$ & $3.000$ & $3.000$ & $\mathbf{6.000}$ & $3.000$\tabularnewline
\hline 
$\kappa$ & $0.030$ & $0.030$ & $0.030$ & $\mathbf{0.060}$\tabularnewline
\hline 
$\mu_{E}$ & $5.000$ & $5.000$ & $5.000$ & $5.000$\tabularnewline
\hline 
$\sigma_{E}^{2}$ & $2.500$ & $2.500$ & $2.500$ & $2.500$\tabularnewline
\hline 
$\mu_{I}$ & $1.772$ & $1.772$ & $1.772$ & $1.772$\tabularnewline
\hline 
$\sigma_{I}^{2}$ & $0.858$ & $0.858$ & $0.858$ & $0.858$\tabularnewline
\hline 
\end{tabular}
\par\end{center}

	\caption{Table of the parameters used in the generation of the simulated data-sets used in Section \ref{sec:sim-study}. \label{tab:simdataparamentersfordatasets}}
\end{table}

For each simulated epidemic, each test was applied using 3 different observation windows corresponding to the intervals up to which 100\%, 70\% or 40\% of the population was observed to be infected.  The likelihood-based tests only allow for estimation of the posterior expectation of an imputed p-value, and we therefore
use the posterior expectation as the summary measure of evidence for all the tests (even though the full posterior can be explored for the 
ILR test).

To each simulated data set $y$ we fit two separate misspecified models with isotropic kernel functions:
\begin{eqnarray*}
K({\bf x}, {\bf y}, \kappa) &= &(1 + |\mathbf{x}-\mathbf{y}|^\kappa)^{-1};\\
K({\bf x}, {\bf y}, \kappa) &= &\exp(-\kappa |\mathbf{x}-\mathbf{y}|^2),\\
\end{eqnarray*}
where $d$ denotes the Euclidean distance between $\bf x$ and $\bf y$. In the former case, infective challenge decreases
according to a power-law, while in the latter case the Gaussian kernel is exponentially bounded. Informally, we may consider the 
first kernel to represent a more severe degree of misspecification, in comparison to the real exponential kernel, than the 
second one. We may anticipate that tests should find more evidence against the assumptions when the power-law kernel is fitted. The fitted model, whose adequacy is to be tested, will be referred to as $M_0$.

The simulated data are generated in all data-sets from an exponential kernel. This model is referred to as $M_1$, and will be the model that $M_0$ is compared against in the LLR tests. This exponential kernel is given by:

\begin{eqnarray*}
	K({\bf x}, {\bf y}, \kappa) &= &\exp(-\kappa |\mathbf{x}-\mathbf{y}|).\\
\end{eqnarray*}

In all cases we use non-informative prior distributions for the parameters in the fitted model as follows: An $\mathrm{Unif}(0,M)$ uniform prior was used for $\alpha,\mu_{E},\sigma_{E}^{2},\mu_{E},\sigma_{E}^{2}$,
where $M\approx1.7\times10^{308}$ is the computer limit for double
precision floating-point numbers in C++. 

The prior distributions used for the other parameters were:

\begin{align*}
\beta & \sim \mathrm{Gamma}(\mu=1,\sigma^{2}=100)\\
\kappa & \sim\mathrm{Gamma}(\mu=1,\sigma^{2}=100)
\end{align*}

\begin{figure}
\adjustbox{max height=\dimexpr\textheight-5.5cm\relax,max width=\textwidth}{\includegraphics{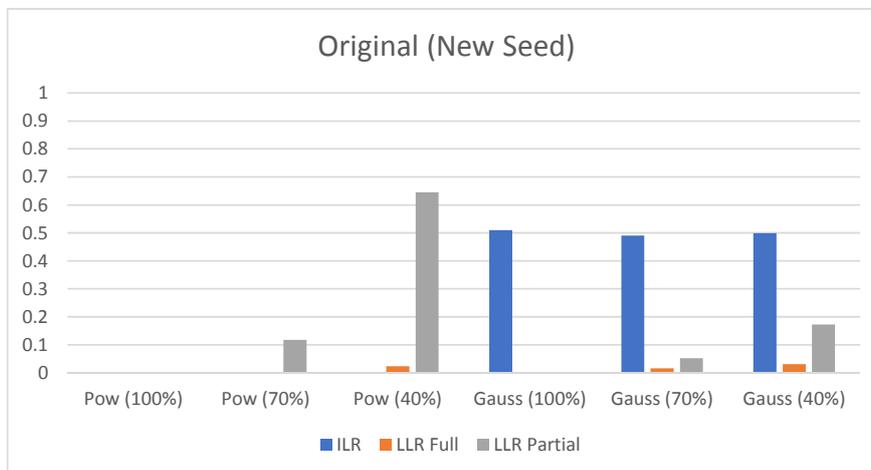}}\caption{Comparison of Latent Likelihood Ratio (LLR) test to Infection Link
Residuals test: Bar chart of the expected posterior $p$-values obtain
for the data set generated with the original parameters, but with
a new random seed for the coordinates of the hosts, where "Pow''
denotes that a power law kernel was fitted and "Gauss'' denotes that a Gaussian
kernel was fitted. The simulated data were observed up to the time
such that a set percentage of the population became infectious. This
percentage is in brackets.\label{fig:barchart-norg}}
\end{figure}

\begin{figure}
\adjustbox{max height=\dimexpr\textheight-5.5cm\relax,max width=\textwidth}{\includegraphics{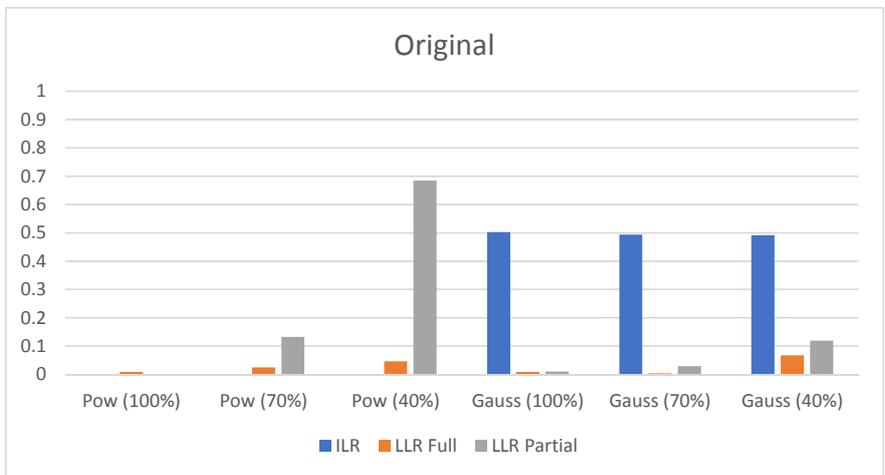}}\caption{Comparison of Latent Likelihood Ratio (LLR) test to Infection Link
Residuals test: Bar chart of the expected posterior $p$-values obtain
for the data set generated with the original parameters, where "Pow''
denotes that a power law kernel was fitted and "Gauss'' denotes that a Gaussian
kernel was fitted. The simulated data were observed up to the time
such that a set percentage of the population became infectious. This
percentage is in brackets.\label{fig:barchart-org}}
\end{figure}

\begin{figure}
\adjustbox{max height=\dimexpr\textheight-5.5cm\relax,max width=\textwidth}{\includegraphics{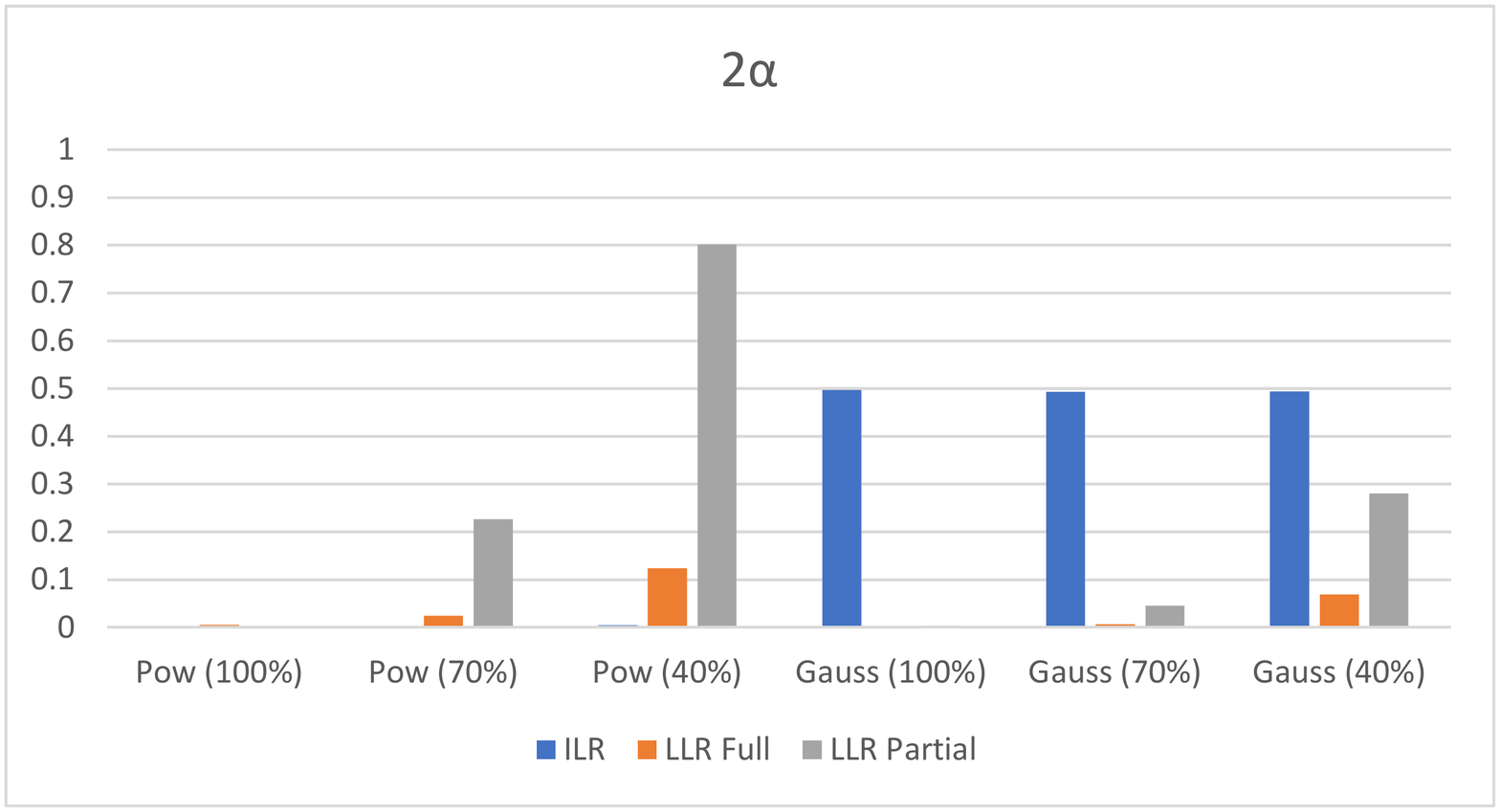}}\caption{Comparison of Latent Likelihood Ratio (LLR) test to Infection Link
Residuals test: Bar chart of the expected posterior $p$-values obtain
for the data set $\alpha\times2$, where "Pow'' denotes that a power
law kernel was fitted and "Gauss'' denotes that a Gaussian kernel was
fitted. The simulated data were observed up to the time such that a
set percentage of the population became infectious. This percentage
is in brackets.\label{fig:barchart-2alpha}}
\end{figure}

\begin{figure}
\adjustbox{max height=\dimexpr\textheight-5.5cm\relax,max width=\textwidth}{\includegraphics{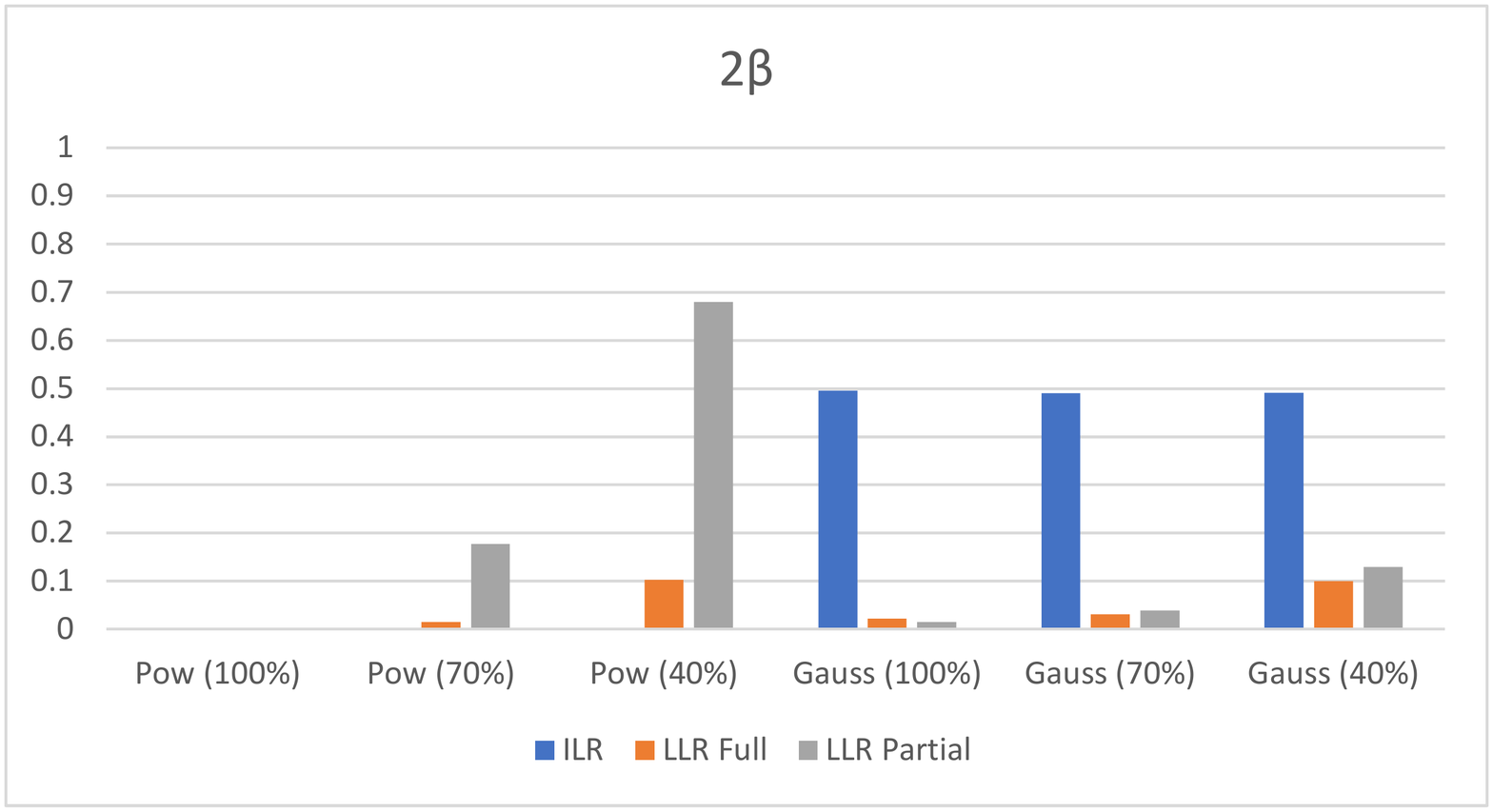}}\caption{Comparison of Latent Likelihood Ratio (LLR) test to Infection Link
Residuals test: Bar chart of the expected posterior $p$-values obtain
for the data set $\beta\times2$, where "Pow'' denotes that a power
law kernel was fitted and "Gauss'' denotes that a Gaussian kernel was
fitted. The simulated data were observed up to the time such that a
set percentage of the population became infectious. This percentage
is in brackets.\label{fig:barchart-2beta}}
\end{figure}

\begin{figure}
\adjustbox{max height=\dimexpr\textheight-5.5cm\relax,max width=\textwidth}{\includegraphics{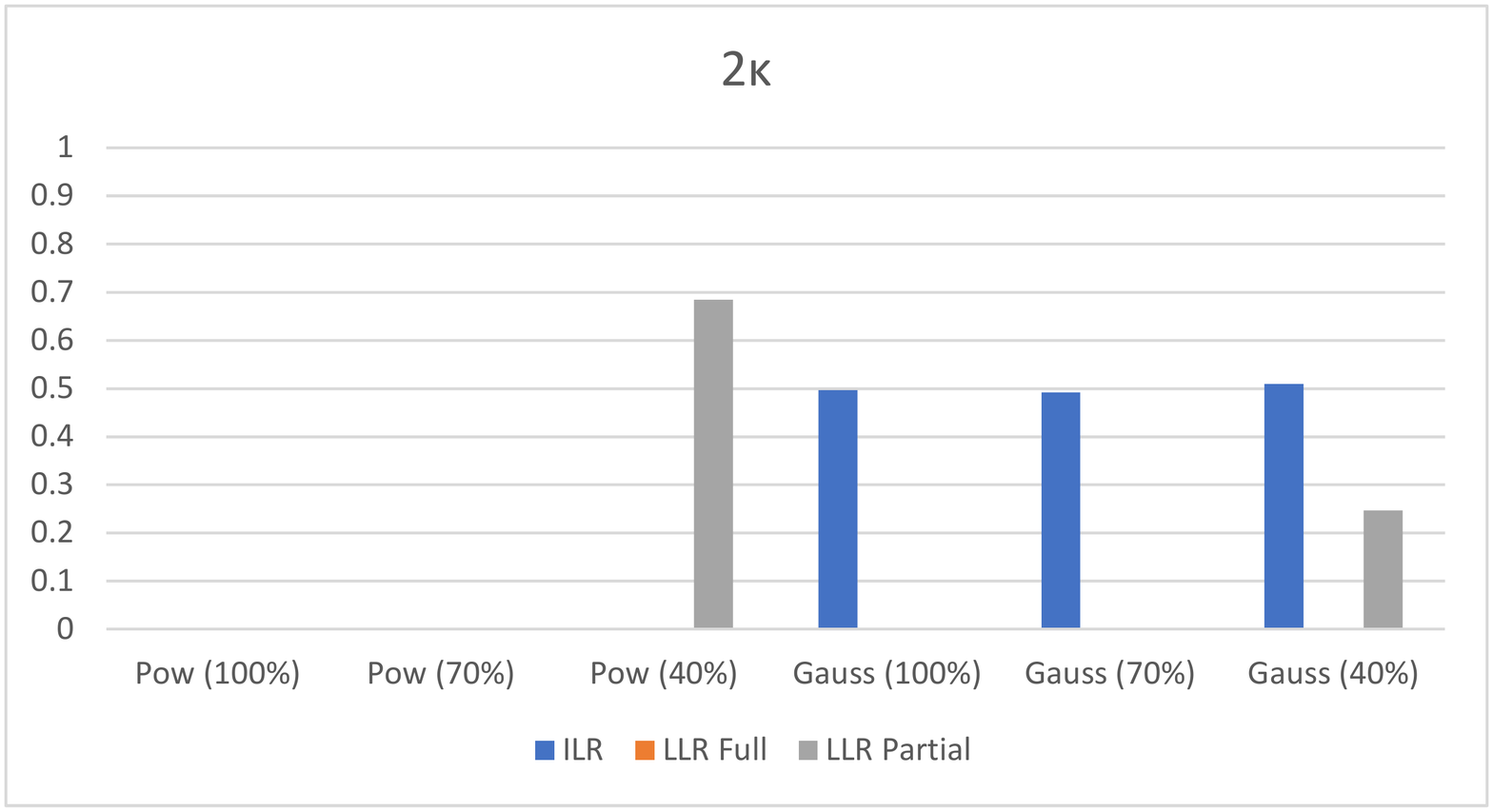}}\caption{Comparison of Latent Likelihood Ratio (LLR) test to Infection Link
Residuals test: Bar chart of the expected posterior $p$-values obtain
for the data set $\kappa\times2$, where "Pow'' denotes that a power
law kernel was fitted and "Gauss'' denotes that a Gaussian kernel was
fitted. The simulated data were observed up to the time such that a
set percentage of the population became infectious. This percentage
is in brackets.\label{fig:barchart-2kappa}}
\end{figure}

The results of the simulation study are presented in Table \ref{tab:Comparison-of-Likelihood-verification} and in Figures \ref{fig:barchart-norg}, \ref{fig:barchart-org},\ref{fig:barchart-2alpha},\ref{fig:barchart-2beta} and \ref{fig:barchart-2kappa}. Some obvious trends can be seen.

\begin{table}
	\adjustbox{max height=\dimexpr\textheight-5.5cm\relax,max width=\textwidth}{\begin{tabular}{|c|c|c|c|c|c|}
		\hline 
		Data-set & $M_{0}$ & {\small{}}\begin{tabular}{c}
			{\small{}Total \%}\tabularnewline
			{\small{}Infections Observed}\tabularnewline
		\end{tabular} & ILR {\small{}$\hat{E(p)}$} & LLR (Full) {\small{}$\hat{E(p)}$} & LLR (Partial) {\small{}$\hat{E(p)}$}\tabularnewline
		\hline 
		\hline 
		$\alpha\times2$ & {\small{}$\left(1+d^{\kappa}\right)^{-1}$} & 100 & 0.0000243 & 0.005319 & 0.0000000\tabularnewline
		\hline 
		$\alpha\times2$ & {\small{}$\left(1+d^{\kappa}\right)^{-1}$} & 70 & 0.0002571 & 0.02473 & 0.2269000\tabularnewline
		\hline 
		$\alpha\times2$ & {\small{}$\left(1+d^{\kappa}\right)^{-1}$} & 40 & 0.0042040 & 0.1242 & 0.8014000\tabularnewline
		\hline 
		$\alpha\times2$ & {\small{}$\exp\left\{ -\kappa d^{2}\right\} $} & 100 & 0.4966585 & 0.0006974 & 0.0038500\tabularnewline
		\hline 
		$\alpha\times2$ & {\small{}$\exp\left\{ -\kappa d^{2}\right\} $} & 70 & 0.4929907 & 0.006932 & 0.0461200\tabularnewline
		\hline 
		$\alpha\times2$ & {\small{}$\exp\left\{ -\kappa d^{2}\right\} $} & 40 & 0.4937340 & 0.06909 & 0.2801000\tabularnewline
		\hline 
		\hline 
		$\beta\times2$ & {\small{}$\left(1+d^{\kappa}\right)^{-1}$} & 100 & 0.0000006 & 0.0000 & 0.0000000\tabularnewline
		\hline 
		$\beta\times2$ & {\small{}$\left(1+d^{\kappa}\right)^{-1}$} & 70 & 0.0000013 & 0.01566 & 0.1771000\tabularnewline
		\hline 
		$\beta\times2$ & {\small{}$\left(1+d^{\kappa}\right)^{-1}$} & 40 & 0.0001413 & 0.1031 & 0.6801000\tabularnewline
		\hline 
		$\beta\times2$ & {\small{}$\exp\left\{ -\kappa d^{2}\right\} $} & 100 & 0.4963660 & 0.02189 & 0.0157500\tabularnewline
		\hline 
		$\beta\times2$ & {\small{}$\exp\left\{ -\kappa d^{2}\right\} $} & 70 & 0.4905312 & 0.03135 & 0.0393000\tabularnewline
		\hline 
		$\beta\times2$ & {\small{}$\exp\left\{ -\kappa d^{2}\right\} $} & 40 & 0.4907014 & 0.1000 & 0.1295000\tabularnewline
		\hline 
		\hline 
		$\kappa\times2$ & {\small{}$\left(1+d^{\kappa}\right)^{-1}$} & 100 & 0.0004014 & 0.0000000 & 0.0000000\tabularnewline
		\hline 
		$\kappa\times2$ & {\small{}$\left(1+d^{\kappa}\right)^{-1}$} & 70 & 0.0002682 & 0.0000000 & 0.0000000\tabularnewline
		\hline 
		$\kappa\times2$ & {\small{}$\left(1+d^{\kappa}\right)^{-1}$} & 40 & 0.0000569 & 0.0000000 & 0.6845000\tabularnewline
		\hline 
		$\kappa\times2$ & {\small{}$\exp\left\{ -\kappa d^{2}\right\} $} & 100 & 0.4970400 & 0.0000000 & 0.0000000\tabularnewline
		\hline 
		$\kappa\times2$ & {\small{}$\exp\left\{ -\kappa d^{2}\right\} $} & 70 & 0.4920980 & 0.0000000 & 0.0021380\tabularnewline
		\hline 
		$\kappa\times2$ & {\small{}$\exp\left\{ -\kappa d^{2}\right\} $} & 40 & 0.5088031 & 0.0000000 & 0.2474000\tabularnewline
		\hline 
		\hline 
		Original & {\small{}$\left(1+d^{\kappa}\right)^{-1}$} & 100 & 0.0000013 & 0.009208 & 0.0000000\tabularnewline
		\hline 
		Original & {\small{}$\left(1+d^{\kappa}\right)^{-1}$} & 70 & 0.0000011 & 0.02533 & 0.1325000\tabularnewline
		\hline 
		Original & {\small{}$\left(1+d^{\kappa}\right)^{-1}$} & 40 & 0.0000569 & 0.04713 & 0.6845000\tabularnewline
		\hline 
		Original & {\small{}$\exp\left\{ -\kappa d^{2}\right\} $} & 100 & 0.5026800 & 0.009208 & 0.0108500\tabularnewline
		\hline 
		Original & {\small{}$\exp\left\{ -\kappa d^{2}\right\} $} & 70 & 0.4943048 & 0.004225 & 0.0294100\tabularnewline
		\hline 
		Original & {\small{}$\exp\left\{ -\kappa d^{2}\right\} $} & 40 & 0.4920137 & 0.06743 & 0.1191000\tabularnewline
		\hline 
		\hline 
		Original (New Seed) & {\small{}$\left(1+d^{\kappa}\right)^{-1}$} & 100 & 0.0000026 & 0.0000 & 0.0000000\tabularnewline
		\hline 
		Original (New Seed) & {\small{}$\left(1+d^{\kappa}\right)^{-1}$} & 70 & 0.0000240 & 0.002046 & 0.1174000\tabularnewline
		\hline 
		Original (New Seed) & {\small{}$\left(1+d^{\kappa}\right)^{-1}$} & 40 & 0.0009413 & 0.02326 & 0.6451000\tabularnewline
		\hline 
		Original (New Seed) & {\small{}$\exp\left\{ -\kappa d^{2}\right\} $} & 100 & 0.5100904 & 0.0007158 & 0.0005900\tabularnewline
		\hline 
		Original (New Seed) & {\small{}$\exp\left\{ -\kappa d^{2}\right\} $} & 70 & 0.4910087 & 0.01579 & 0.05212\tabularnewline
		\hline 
		Original (New Seed) & {\small{}$\exp\left\{ -\kappa d^{2}\right\} $} & 40 & 0.4991936 & 0.03086 & 0.1722\tabularnewline
		\hline 
\end{tabular}}
	
	\caption{Comparison of Latent Likelihood Ratio (LLR) test to Infection Link
		Residuals test: data-set, $M_0$ tested and estimated expected
		$p$-values from the infection link residuals test, LLR (full likelihood)
		and LLR (partial likelihood)\label{tab:Comparison-of-Likelihood-verification}}
\end{table}


\begin{itemize}
\item
The ILR based test consistently finds very strong evidence against the model when the power-law kernel is wrongly fitted.  However, 
no evidence emerges when the exponentially bounded, Gaussian kernel is fitted to the observations, with the mean p-value being close to 
0.5. This suggests that the ILR test may be insensitive to misspecification if the degree of discrepancy is modest.
When the data are simulated using the larger value of $\kappa$ (so that secondary infection tends to occur over short range) 
and the power-law kernel is fitted, the evidence against the assumptions
is strongest when only 40\% of the hosts are infected.  This may be due to the short-range secondary infection being most apparent 
during the early stages of the epidemic where the pattern of infection is clearly formed from isolated foci (caused by primary infection) surrounded by
clustered secondary infections.  As a result, residuals from the early stage of the epidemic (when the potential choice of exposure locations is widest)
may display the greatest evidence against the assumed model and inclusion 
of residuals from later in the epidemic may serve to dilute this evidence.  Simulated epidemics are presented in electronic Appendix 3 to illustrate this point.
\item 
In contrast, the LLRT based on the full likelihood elicits some evidence against 
the assumptions in all cases, including when the Gaussian kernel is fitted. Moreover, in all cases 
the strength of the evidence as measured by the expected p-value increases as the observation duration increases (and thus the percentage of observed infections).
\item
The performance of the LLRT that uses the partial likelihood seems variable. It detects evidence in 
cases where the observation duration is long, but appears to degrade as less of the epidemic is observed. Therefore, the full LLR test may be a more robust approach. 
\item
When we focus on the case where only 40\% of the population is infected, we see that the
ILR test typically provides most evidence when the power-law kernel is assumed.  The full likelihood approach performs best
when the Gaussian kernel is assumed. This last observation may illustrate the phenomenon of reinforcement discussed earlier; the imputed 
data reinforce the `wrong' assumptions of the fitted model and undermine to some extent the capacity of the 
imputed likelihood ratio test to find evidence against the assumed model when it is poorly specified.
\end{itemize}

In order to determine whether the tests produce false positives when the fitted kernel is the same as the actual kernel, an extra set of computer runs were performed in which an exponential kernel model was fitted to the simulated data generated from a model with an exponential kernel.  The results can be found in Table \ref{tab:rightwayround}.

\begin{table}
	\adjustbox{max height=\dimexpr\textheight-5.5cm\relax,max width=\textwidth}{\begin{tabular}{|c|c|c||c|c|c|}
			\hline 
			Data-set & $M_{A}$ & {\small{}}\begin{tabular}{c}
				{\small{}Total \%}\tabularnewline
				{\small{}Population}\tabularnewline
				{\small{} Infectious}\tabularnewline
			\end{tabular} & ILR {\small{}$\hat{E(p)}$} & LLR (Full) {\small{}$\hat{E(p)}$} & LLR (Partial) {\small{}$\hat{E(p)}$}\tabularnewline
			\hline 
			\hline 
			Original & {\small{}$\exp\left\{ -\kappa d^{2}\right\} $} & 100 & 0.502260 & 0.5410 & 0.649163\tabularnewline
			\hline 
			Original & {\small{}$\exp\left\{ -\kappa d^{2}\right\} $} & 70 & 0.5032322 & 0.5910 & 0.6824\tabularnewline
			\hline 
			Original & {\small{}$\exp\left\{ -\kappa d^{2}\right\} $} & 40 & 0.4898694 & 0.8043 & 0.7718\tabularnewline
			\hline 
	\end{tabular}}
	
	\caption{Comparison of Latent Likelihood Ratio (LLR) test to Infection Link
		Residuals test: data-set, alternative model tested and estimated expected
		$p$-values from the infection link residuals test, LLR (full likelihood)
		and LLR (partial likelihood)\label{tab:rightwayround}}
\end{table}

The values of $\hat{E(p)}$ in Table \ref{tab:rightwayround} show
that performing a test with $M_{0}$ being the model with the true
kernel and $M_{A}$ being an alternative model with a similar spatial
kernel, all tests do not detect discrepancy between the fitted model
and the alternative model, thus showing the test do not have a propensity to produce false positives.

\section{Conclusions \& discussion}

In this paper we have investigated methods for assessing and comparing spatio-temporal stochastic epidemic models - particularly 
with regard to the specification of the spatial kernel function.  We have focussed on techniques that avoid the need to increase the complexity
of the fitted model (e.g. by specifying more highly parametrised kernel functions). Rather, the methods that we consider
can be implemented as relatively straightforward addenda to a Bayesian analysis where the model criticism is achieved by
embedding classical testing methods within the Bayesian analysis - in the same spirit as posterior predictive checking.
In particular, we have compared the ability of the infection-link residuals 
introduced by Lau at al \cite{Lau14} to detect kernel misspecification
with that of tests based on latent likelihood ratio tests. The simulation study uses data in which the transition into the I and R states are observed, but can be easily adapted to snapshot data, data with under-reporting and other forms of data censoring \cite{Gamado2013}, where epidemic model selection is often hindered by computational complexity.   The results demonstrate that
the former approach performs well when the degree of model misspecification in high - that is when
a power-law kernel is assumed when the true kernel is exponential - but is unable to detect evidence 
when the true and assumed kernels are qualitatively more similar.  On the other hand, a test based on
a full latent likelihood is able to elicit evidence of the more subtle misspecification.

The results point to an interesting phenomenon regarding the use of classical tests applied to imputed processes.
Since the additional data are imputed using the misspecified model, it need not follow that basing the testing on
more data leads to more power. In certain cases, the ILR methodology applied to the emergent phase of the epidemic only
provides more evidence of discrepancy than when the full trajectory is used. This in turns leads to the 
notion of how best to design a latent experiment.  How can one use prior belief on model parameters to 
predict (before data are considered) which form of latent test will be best able to detect a suspected mode
of misspecification? Answering this question is a challenge which we seek to address in ongoing work. Nevertheless,
we suggest that the techniques presented in this paper can offer readily implementable ways of checking model 
assumptions while avoiding the complexities and instabilities associated with a purely Bayesian approach.

\section*{Acknowledgements}
\noindent David Thong thanks the EPSRC for providing funding for the PhD project from which this research arises.

\clearpage
\bibliographystyle{spmpsci}      
 {}

\end{document}